\documentclass{vgtc}

\ifpdf%
  \pdfoutput=1\relax
  \usepackage{graphicx}
  \DeclareGraphicsExtensions{.pdf,.png,.jpg,.jpeg}
\else%
  \usepackage{graphicx}
  \DeclareGraphicsExtensions{.eps}
\fi%

\graphicspath{{figures/}{pictures/}{images/}{./}}

\usepackage{microtype}
\PassOptionsToPackage{warn}{textcomp}
\usepackage{textcomp}
\usepackage{mathptmx}
\usepackage{times}

\usepackage{cite}
\usepackage{tabu}
\usepackage{booktabs}
\usepackage{caption}

\onlineid{0}

\vgtccategory{Research}

\vgtcinsertpkg

\title{RayPC: Interactive Ray Tracing Meets Parallel Coordinates}

\author{Jonathan Fritsch\thanks{e-mail: jonathan.fritsch@dlr.de} %
\and Markus Flatken %
\and Simon Schneegans %
\and Andreas Gerndt %
\and Ana-Catalina Plesa
\and Christian Hüttig
} %
\affiliation{\scriptsize German Aerospace Center (DLR)}

\teaser{
  \includegraphics[width=\linewidth]{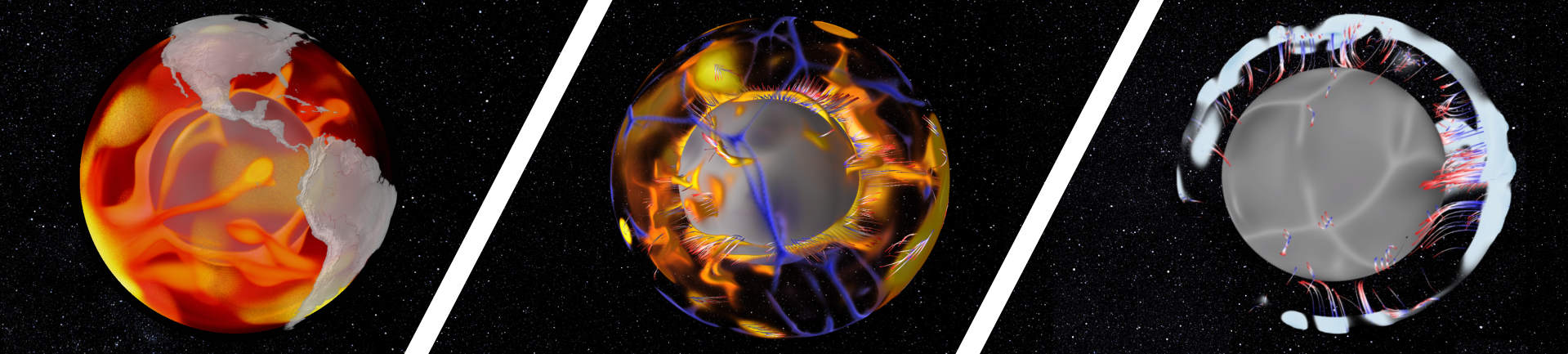}
  \caption{Interactive visualisation and analysis of mantle flow in the Earth's interior: hot mantle plumes (left image), mantle plumes with associated pathlines and cold downwellings shown in blue color (center image), stagnated cold material (light blue) and pathlines indicating sinking of downwellings in the lower mantle (right image).}

}

\abstract{
Large-scale numerical simulations of planetary interiors require dedicated visualization algorithms that are able to efficiently extract a large amount of information in an interactive and user-friendly way. Here we present a software framework for the visualization of mantle convection data. This framework combines real-time volume rendering, pathline visualization, and parallel coordinates to explore the fluid dynamics in an interactive way and to identify correlations between various output variables.
}

\renewcommand{\authorfootertext}{\copyright 2021 IEEE\\\tiny Personal use of this material is  permitted. Permission from IEEE must be obtained for all other uses, in any current or future media, including reprinting/republishing this material for advertising or promotional purposes, creating new collective works, for resale or redistribution to servers or lists, or reuse of any copyrighted component of this work in other works.\\\texttt{https://virtual.ieeevis.org/year/2021/paper\_a-sciviscontest-1006.html}}

\begin{document}

\firstsection{Motivation}

\maketitle

\makeatletter
\renewcommand{\thefootnote}{}%
\footnotetext[0]{
	\begin{flushleft}
		\vskip -40pt
		\begin{list}{\textbullet}{
				\setlength{\partopsep}{0pt}
				\setlength{\topsep}{0pt}
				\setlength{\itemsep}{-2pt}
				\setlength{\itemindent}{-4pt}
				\setlength{\leftmargin}{12pt}}
			\authorfootertext
		\end{list}
	\end{flushleft}
}%
\renewcommand{\thefootnote}{\arabic{footnote}}
\makeatother

Over the past decades, with the increase of processing power and available memory, global-scale computational models have become state-of-the-art to investigate the complex physical processes in the interior of the Earth and other rocky planets in the Solar System, e.g., \cite{crameri2014,plesa2018,shahnas2017}.
Numerical models that employ a high spatial resolution are widely used to link the evolution of mantle flow to the formation and evolution of surface structures, such as volcanoes and tectonic plates. 
Depending on the research question, the size of each snapshot file, which is typically written during a simulation and contains output quantities such as the temperature field, velocity, thermal conductivity and expansivity etc., can increase considerably.
This requires a dedicated visualization software that can handle a large volume of data and present it in a user friendly way.
In particular, such a visualization software needs to be able to quickly react to user input, increasing the level of interactivity, and to provide an intuitive option to identify potential correlations between output quantities.

In this study, we present a new plugin for the open-source software CosmoScout~VR \cite{Schneegans2020}, which can visualize large-scale simulation data of a planet's interior.
In Section~\ref{sec:approach}, we give some implementation details.
We used the plugin to fulfill the five given tasks and present the results in Section~\ref{sec:results} and conclude with a discussion and plans for future work in Section~\ref{sec:discussion}.  

\section{Approach} \label{sec:approach}
\begin{figure}[t]
  \centering
  \includegraphics[width=\linewidth]{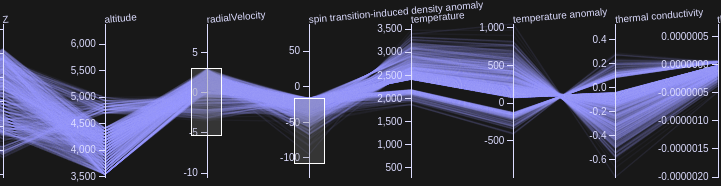}
  \caption{A parallel coordinates plot enables efficient data selection and visual exploration of correlations between variables.}
  \label{fig:parallelcoordinate}
\end{figure}

The idea of our implementation is to use well-known visualization approaches from different domains in one interactive 3D rendering application.
We combine real-time volume rendering and pathline visualization with a parallel coordinates plot which is usually used in the field of information visualization.
Figure~\ref{fig:parallelcoordinate} shows an example of the parallel coordinates.
This not only allows users to visually detect correlations between variables, but also enables the specification of upper and lower bounds for each variable.
These bounds are then used during rendering to only visualize a subset of the data (\emph{brushing and linking}).
This mechanism enables efficient isolation of structures such as plumes or cold slabs.
Additionally, we display pathlines spawned inside the currently selected subset of the data.
This provides an idea of where the selected material will move to.

In order to achieve interactive volume visualization we use an asynchronous rendering approach.
The frame rate of the OSPRay based ray tracing is decoupled from the main render thread.
In order to hide the varying frame rates and latencies an image-warping technique is used.
To further increase the performance of our volume rendering, we use a level-of-detail data structure and progressive render passes.
All together, this enables explorative data analysis and interactive visualization.

\subsection{Data Preprocessing}

To reduce the memory and I/O requirements, as well as to enhance the data with derived information, we introduce a preprocessing step. 
First, we generate an octree-based level-of-detail data structure with two lower resolution levels.
Each level only contains $1/8^{th}$ the number of cells as the previous level.
Second, we compute additional derived variables (radial velocity and altitude) which help the identification of plumes or slabs at specific heights using the parallel coordinates.
Third, we randomly select up to $100\,000$~cells and store their scalar information in a file.
This is later used to draw the lines of the parallel coordinates.
Lastly, in order to be able to correlate volume structures (plumes or slabs) with the velocity flow field, we generate pathlines seeded at critical points of the temperature anomaly field.
The critical points of each time step were identified by using the TopologyToolkit (TTK)\cite{Tierny_2018} and then traced forward for 10\,time steps (20\,Myrs).
This enables the visualization of dynamic behavior in a static image as it depicts where the selected material will move to.

\subsection{Rendering}

Our volume rendering is based on the open-source ray tracing engine OSPRay from Intel\footnote{https://www.ospray.org/}.
We implemented an asynchronous rendering approach, so that the application frame rate does not depend on the frame rate of the volume rendering.
The missing frames are interpolated by using an image-warping technique.
This guarantees stable and high frame rates which is key for an immersive experience in virtual reality. 
While the virtual camera is stationary, the previously rendered image is enhanced by running additional render passes.
This increases the number of samples, that are available per pixel and reduces the image noise inherent to ray traced images.

For interactive feature selection using the parallel coordinates, the OSPRay renderer was extended to query all relevant variables during ray-marching.
If the queried values for a volume sample do not satisfy the restrictions defined in the parallel coordinates, the sample is discarded and does not contribute to the ray's final color.
As this approach iterates over all restricted variables for each volume sample, there is an overhead in comparison to render without any restrictions, that scales linearly with the number of restricted variables.
However, since the volume data is not modified when changing the restrictions, no acceleration structures have to be rebuild.
This results in a minimal latency between changing the restrictions and displaying the resulting volume.
Also, it allows a more interactive evaluation of different parallel coordinate configurations.
The user interface of our application employs the Chromium Embedded Framework, which lets us use the parcoords-es\footnote{https://github.com/BigFatDog/parcoords-es} library for drawing and interacting with the parallel coordinates.

The pathlines generated in the preprocessing step are rendered with OSPRay, too.
This results in an easier composition with the volume and enables the usage of OSPRay's ambient occlusion to enhance the visual quality of the pathlines.
By default, all pathlines spawned during the last 10 time steps, that started in the current volume selection are displayed.
Optionally, the visible pathlines can be further refined by using another parallel coordinates plot.

\section{Results}\label{sec:results}

We applied our visualization software to address the five given tasks using the data set provided by the Pysklywec Lab of the University of Toronto.
Snapshots of the interactive visualization are provided in the appendix. The accompanying video extensively demonstrates the visualization features.

\noindent \textbf{Task 1:}
Here we show stagnated or diverted cold slabs (descending mantle material) at $\sim660$\,km (upper and lower mantle boundary) depth.
To this end, we use the parallel coordinates and select an altitude range of $\pm 100$\,km around $660$\,km depth, a non-positive radial velocity, suggesting stagnated and sinking material, and a negative temperature anomaly, indicative of cold downwellings. We use the radial velocity as transfer function, to color the areas of stagnated slabs (radial velocity $\sim0$). Sinking slabs can also be seen in the volume rendering since we selected a non-positive range for the radial velocity.

\noindent \textbf{Task 2:}
In this task, we visualize stagnated or diverted cold slabs at $\sim1600$\,km (mid-mantle) depth. As before, we select a negative temperature anomaly for volume rendering. However, this time we did not restrict the depth in order to see where the cold material comes from and where it moves to. We select the spin transition-induced density anomaly as color. A positive density anomaly indicates sinking downwellings, while a negative one shows stagnating downwellings. By selecting the entire altitude range we can observe the behavior of downwellings over the entire mantle (i.e., stagnated at 660-phase transition, stagnated or diverted in the mid-mantle or sinking to the core-mantle boundary).

\noindent \textbf{Task 3:}
Here we identify stagnated or diverted hot plumes (rising hot mantle material) at $\sim1600$\,km depth and their rise to the upper regions of the lower mantle. In the parallel coordinates we select the regions with a positive temperature anomaly and the entire altitude range. This actually allows to observe stagnated and deflected plumes both in the mid-mantle and at 660\,km depth. We distinguish between stagnating and rising plumes by using the spin-transition induced density anomaly as color for volume rendering. A negative density anomaly indicates rising plumes, while a positive one indicates stagnating plumes.

\noindent \textbf{Task 4:}
Stagnated or diverted hot plumes at $\sim660$\,km depth have been already seen in the previous task. An additional way to identify these features is to select a positive temperature anomaly, choose an altitude around $\sim660$\,km, and use the radial velocity for the coloring. This shows stagnating and rising slabs at $\sim660$\,km in a similar fashion as the downwellings in Task 1.

\noindent \textbf{Task 5:}
Correlations between the variables and the flow patterns can be easily identified in our framework through the parallel coordinates. The user can choose various value ranges for simulation variables and selected flow patterns. For instance, a positive thermal expansivity and temperature anomaly can be used to select plumes in the lower mantle. Cold material in the upper mantle can be selected by a negative expansivity and temperature anomaly. More complex correlations can be derived by including additional variables (e.g., spin transition). Additionally, pathlines can help visualize the mantle flow (i.e., if material rises/sinks or glides along boundaries).

\section{Discussion and Future Work}\label{sec:discussion}
In this study, we presented an open-source software framework that offers volume rendering, pathlines and parallel coordinates to analyze large-scale data from geodynamic models.
It provides a high level of interactivity and the ability to easily identify correlations between various output variables. 
By fulfilling the tasks of the contest, we demonstrated that our framework is suitable for the visualization of mantle plumes and downwellings.
However, our approach is not limited to these features and can be used also for many other tasks.
For example, tasks which involve the tracking of material flow rather than stagnating regions can benefit especially from the pathline feature.
In the future, we plan to apply our framework to various tasks, data sets, or applications (e.g., climate data).

\bibliographystyle{abbrv-doi}

\bibliography{template}

\newpage
\appendix

\begin{figure*}
    \centering
    \captionsetup{width=0.85\textwidth}
    \includegraphics[width=0.85\textwidth]{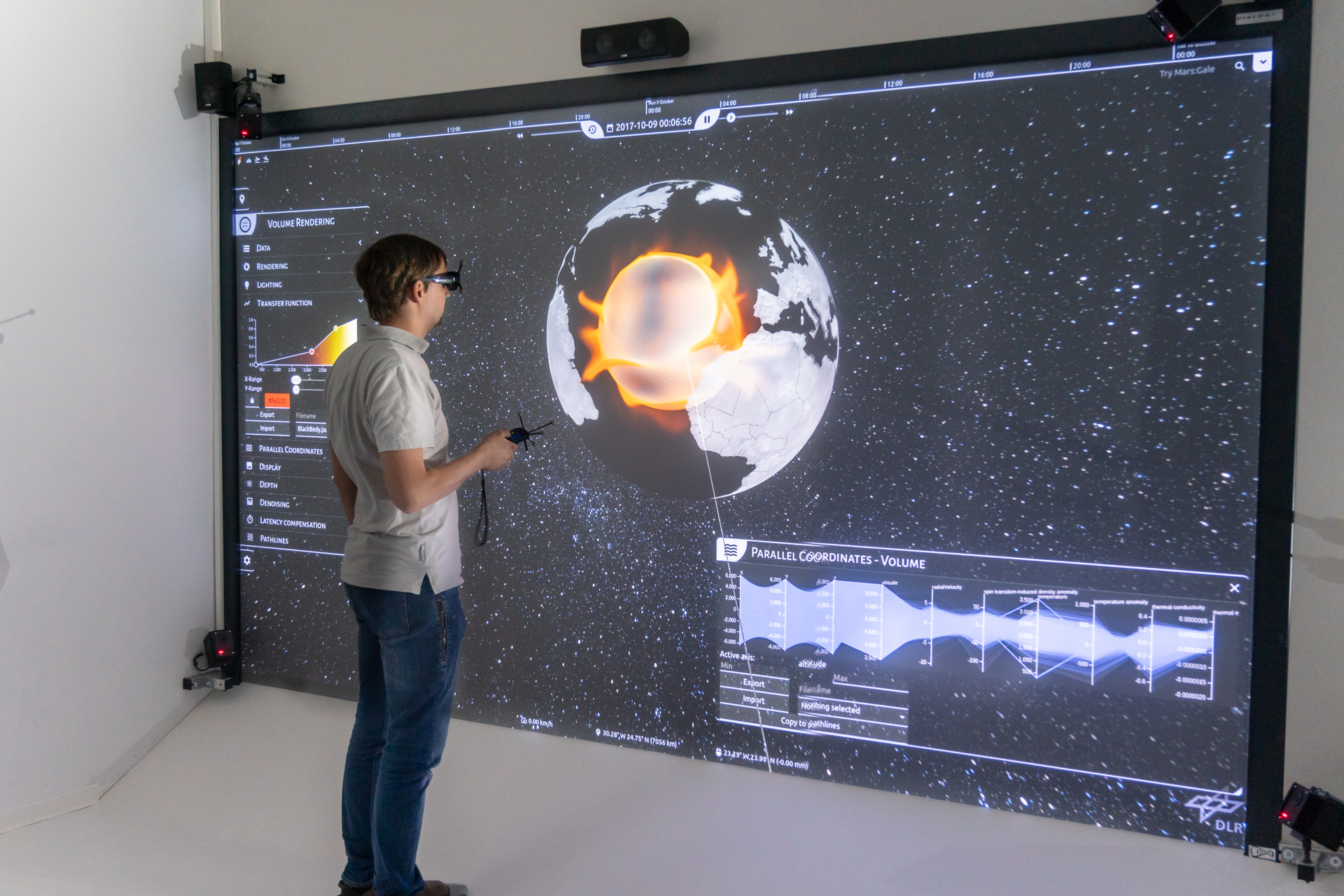}
    \caption{Using the system on a stereoscopic output device with head-tracking significantly improves the understanding of the complex plume geometry.}
\end{figure*}

\begin{figure*}
    \centering
    \captionsetup{width=0.85\textwidth}
    \includegraphics[width=0.85\textwidth]{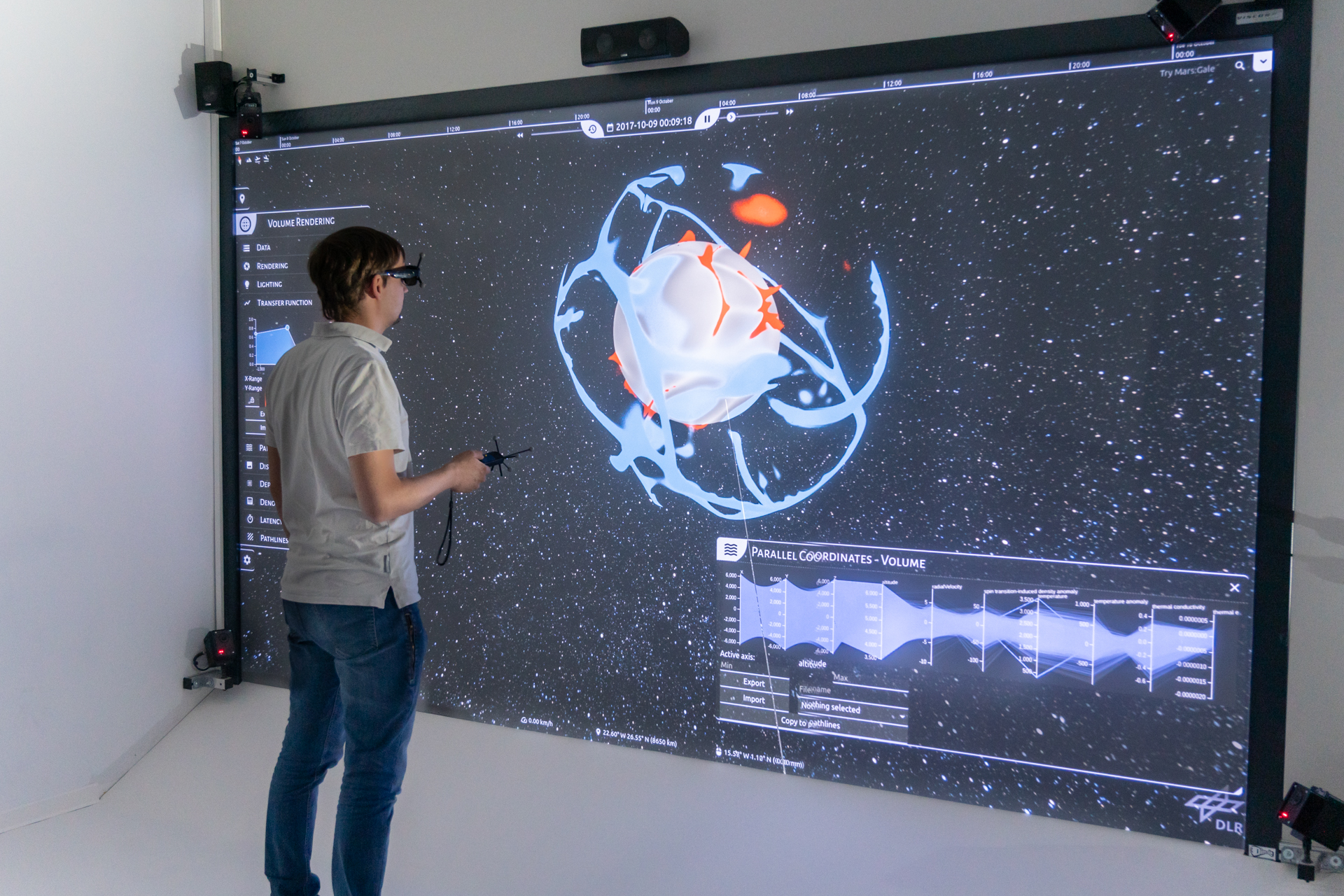}
    \caption{Here we used the transfer function editor to highlight regions of a high (orange) and low (blue) temperature anomaly.}
\end{figure*}

\begin{figure*}
    \centering
    \captionsetup{width=0.85\textwidth}
    \includegraphics[width=0.85\textwidth]{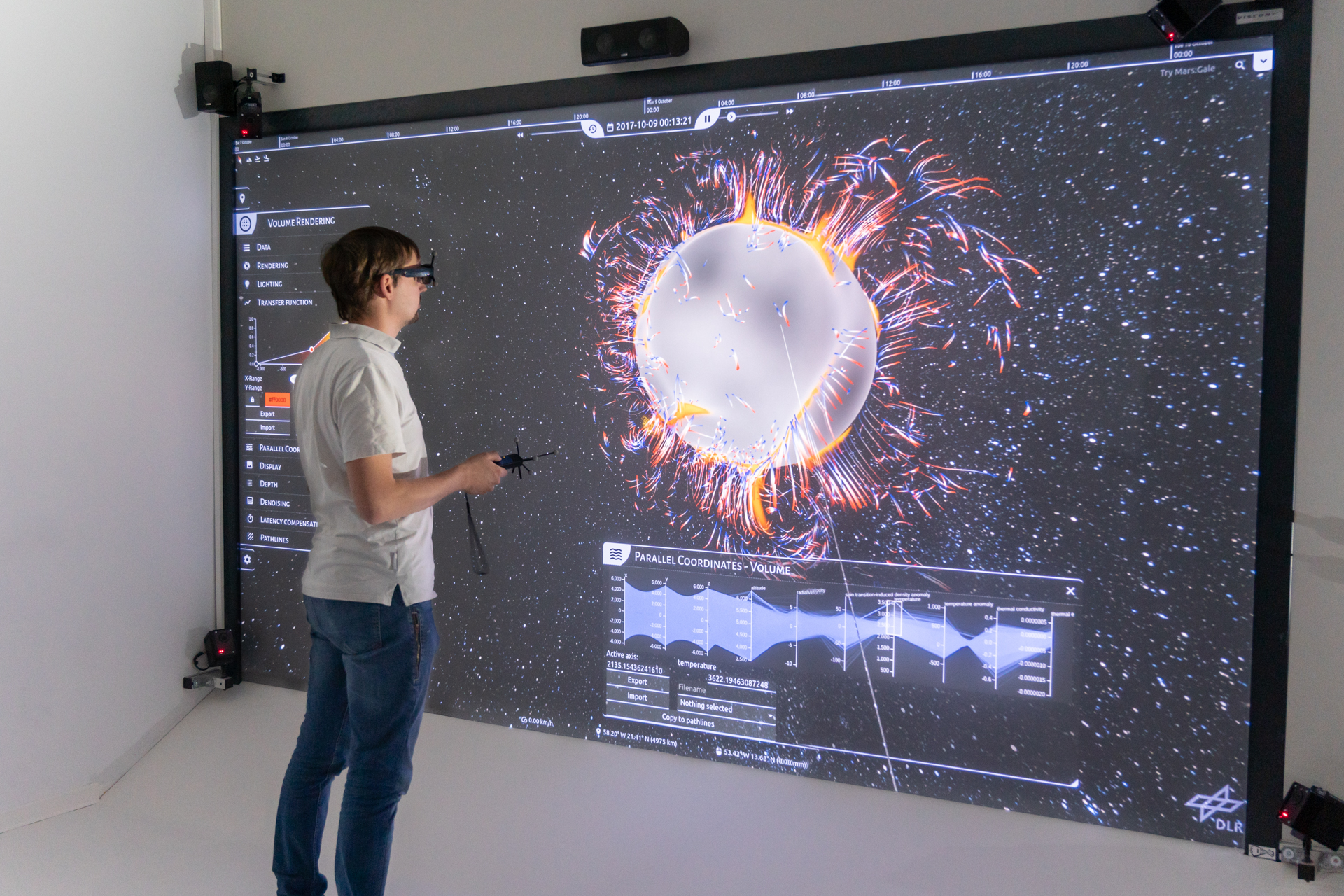}
    \caption{Here we used the parallel coordinates to select regions with a high absolute temperature. The pathlines show where the selected material will move in the future. Convection cells are clearly visible.}
\end{figure*}

\begin{figure*}
    \centering
    \captionsetup{width=0.85\textwidth}
    \includegraphics[width=0.85\textwidth]{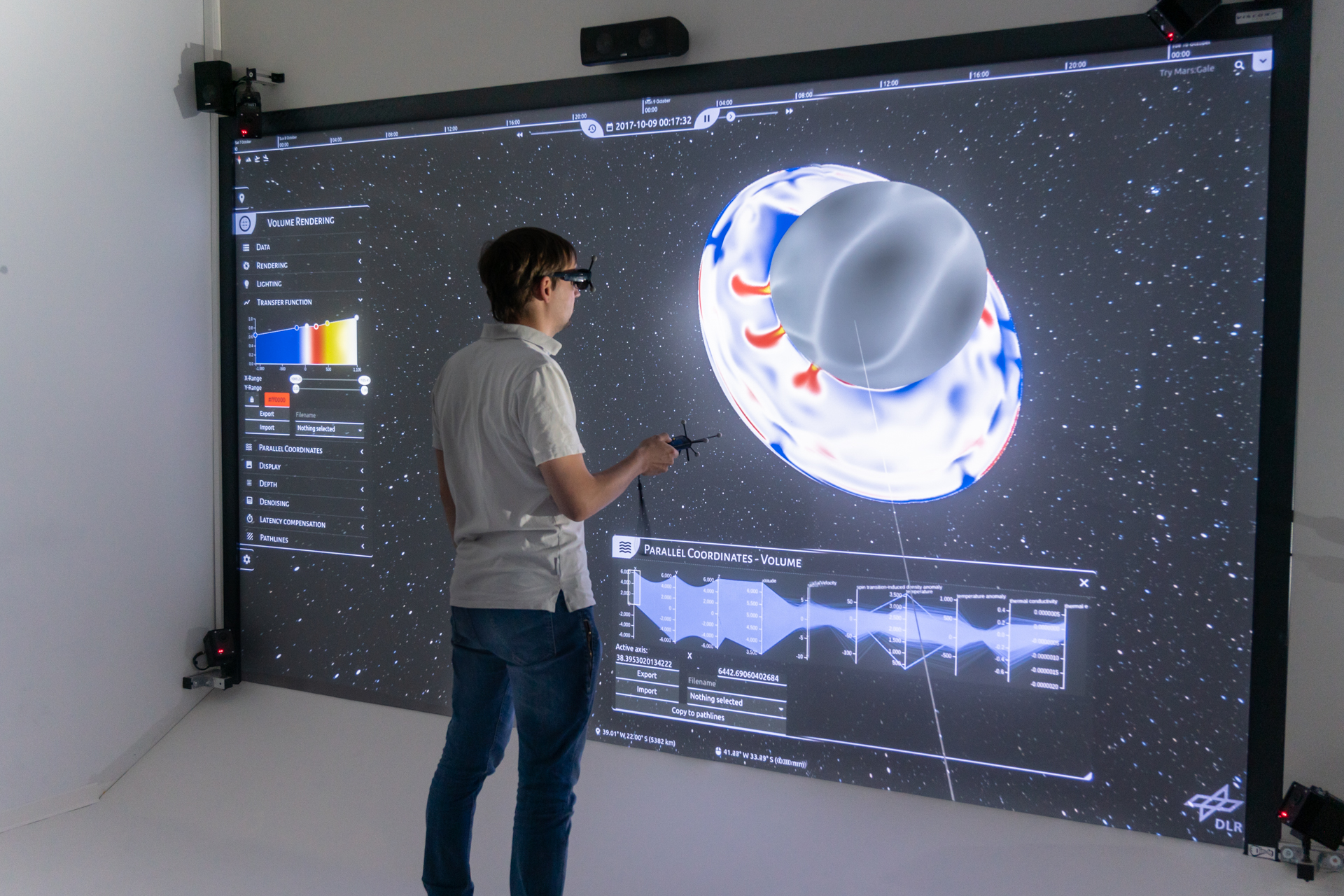}
    \caption{We can use the parallel coordinates not only to limit the volume based on scalars but also based on Cartesian coordinates. This can be used to create cuts through the volume.}
\end{figure*}

\begin{figure*}
    \centering
    \captionsetup{width=0.9\textwidth}
    \includegraphics[width=0.9\linewidth]{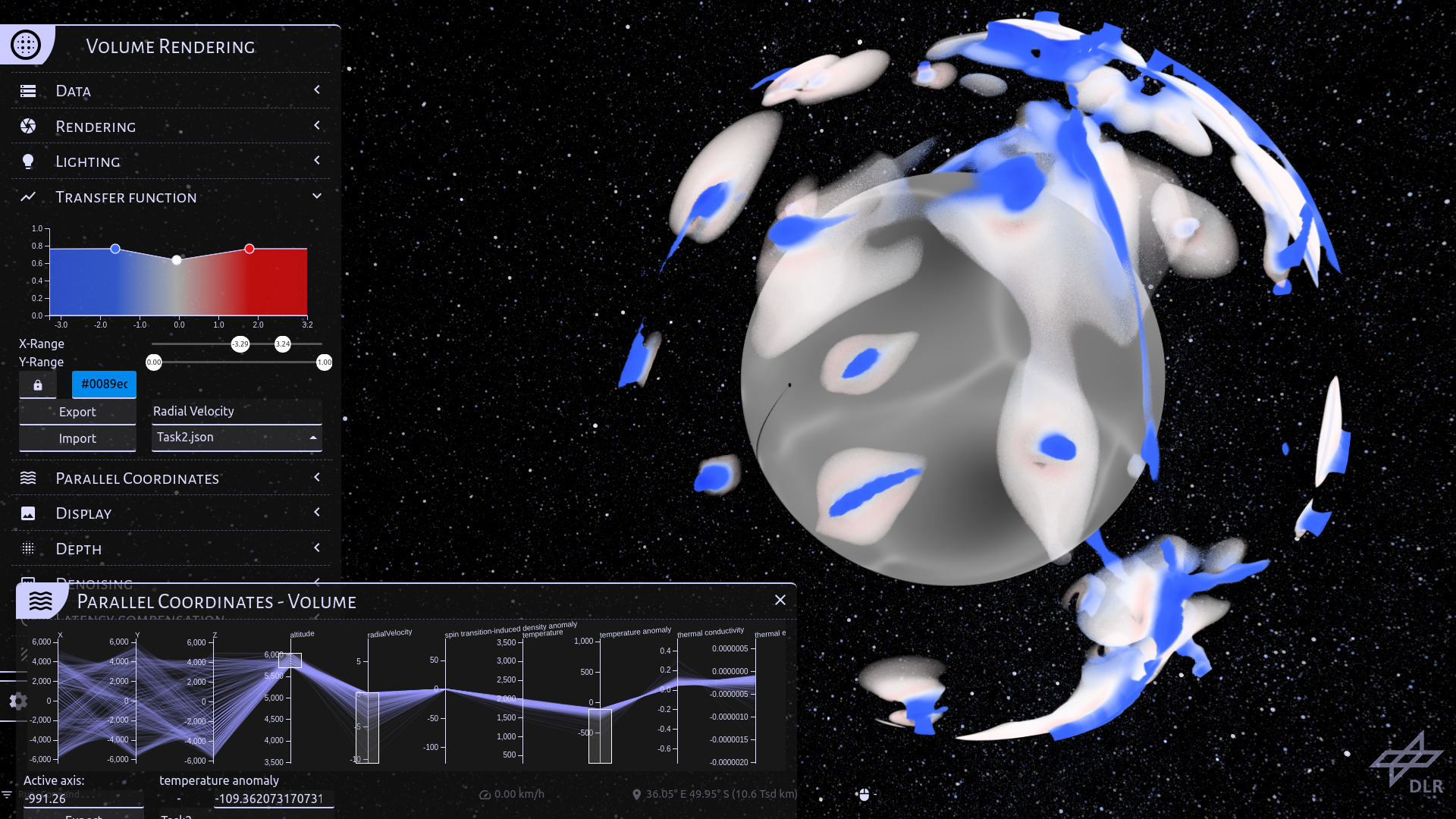}
    \caption{For \textbf{Task 1}, we selected a non-positive radial velocity, a negative temperature anomaly, and a depth range around 660\,km. While the cold slabs are well visible in this image, their behavior is much more apparent while playing the animation. The transfer function uses the radial velocity: Stagnating slabs are shown in white while sinking slabs are colored blue.}
\end{figure*}

\begin{figure*}
    \centering
    \captionsetup{width=0.9\textwidth}
    \includegraphics[width=0.9\linewidth]{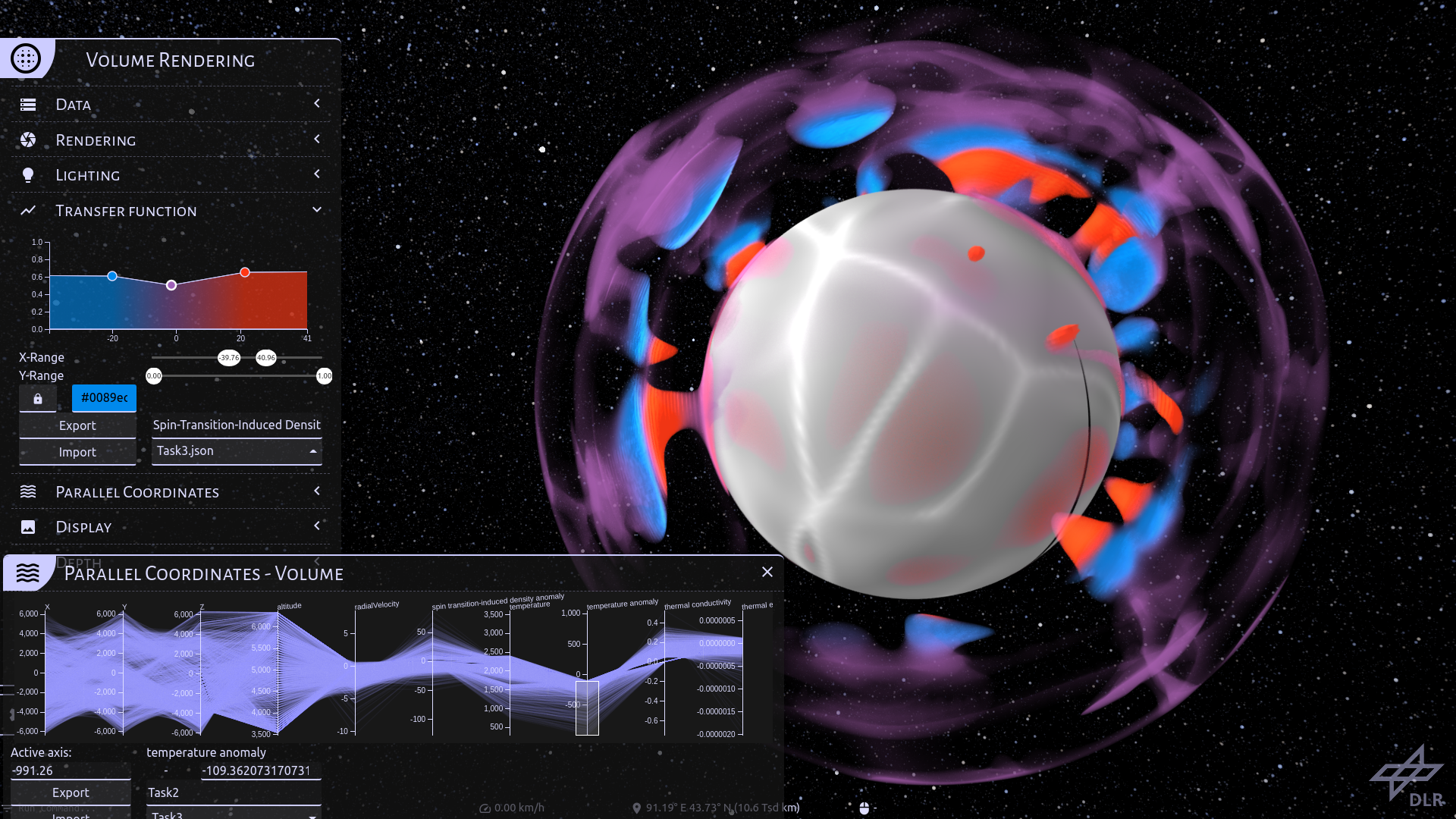}
    \caption{For \textbf{Task 2}, we select a negative temperature anomaly and colorize the downwellings according to the spin transition-induced density anomaly. A negative density anomaly (blue color) causes the downwellings to stagnate and move laterally at mid-mantle depth ($\sim1600$\,km). A positive density anomaly (red color) indicates a downward acceleration of the downwellings. Especially in the upper left area of this time step, we can clearly see how cold material flows downward until it slows down and spreads out as it approaches the iron-spin transition at about 1600\,km depth.}
\end{figure*}

\begin{figure*}
    \centering
    \captionsetup{width=0.9\textwidth}
    \includegraphics[width=0.9\linewidth]{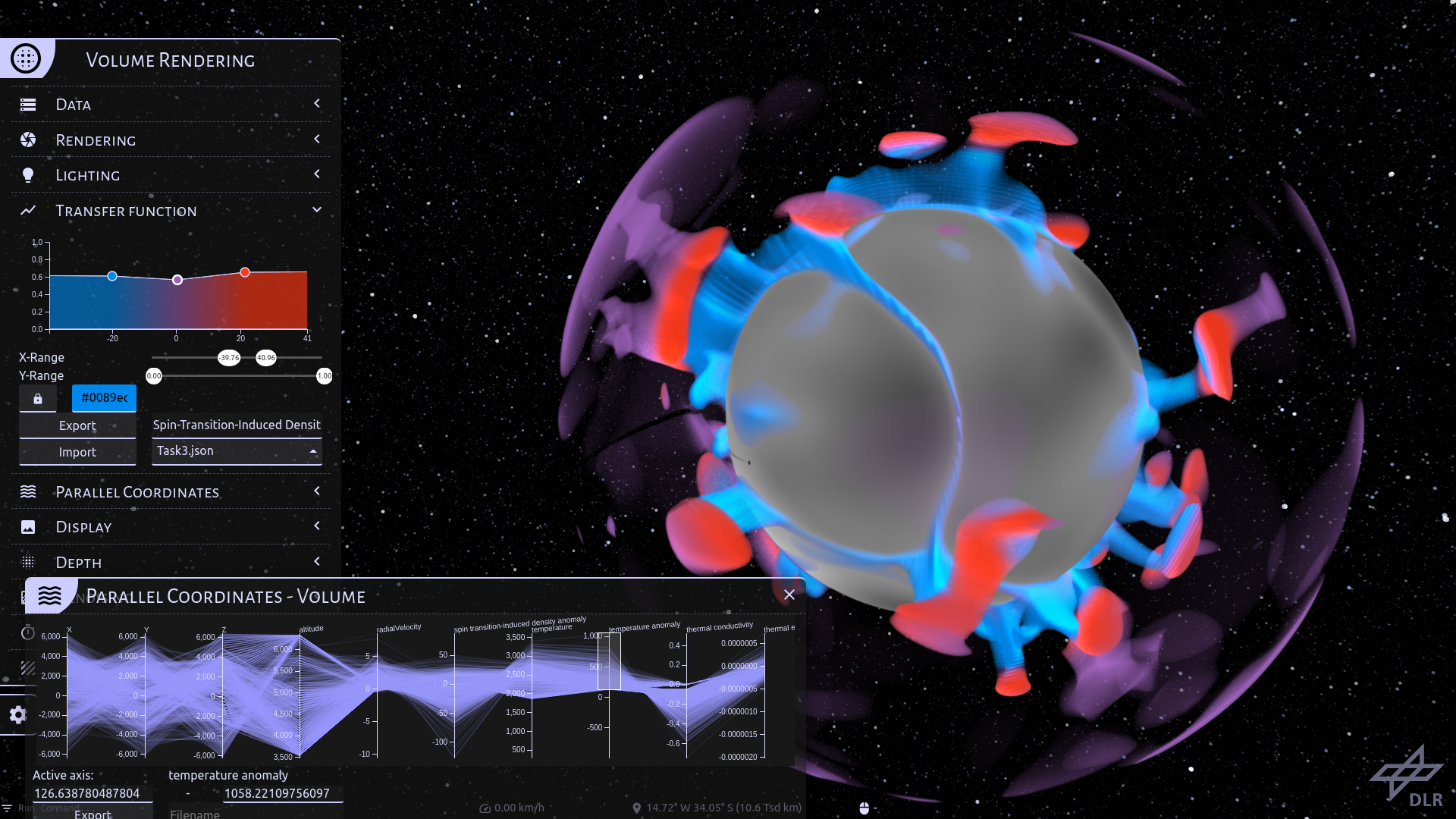}
    \caption{We used a positive temperature anomaly for \textbf{Task 3}. The transfer function is based on the spin transition-induced density anomaly. A negative density anomaly (blue color) indicates rising plumes. A positive density anomaly (red color) causes the plumes to stagnate and move laterally at mid-mantle depth ($\sim1600$\,km). We also show density anomaly values close to zero (purple regions) that are attained at depths shallower than 1500\,km. Here, the endothermic phase transition at 660-km depth is clearly visible, being indicated by the thickening and thinning of plumes in the upper mantle. Some plumes are deflected at around $1600$\,km depth and additionally at the 660-phase boundary (see structure at about three o'clock).}
\end{figure*}

\begin{figure*}
    \centering
    \captionsetup{width=0.9\textwidth}
    \includegraphics[width=0.9\linewidth]{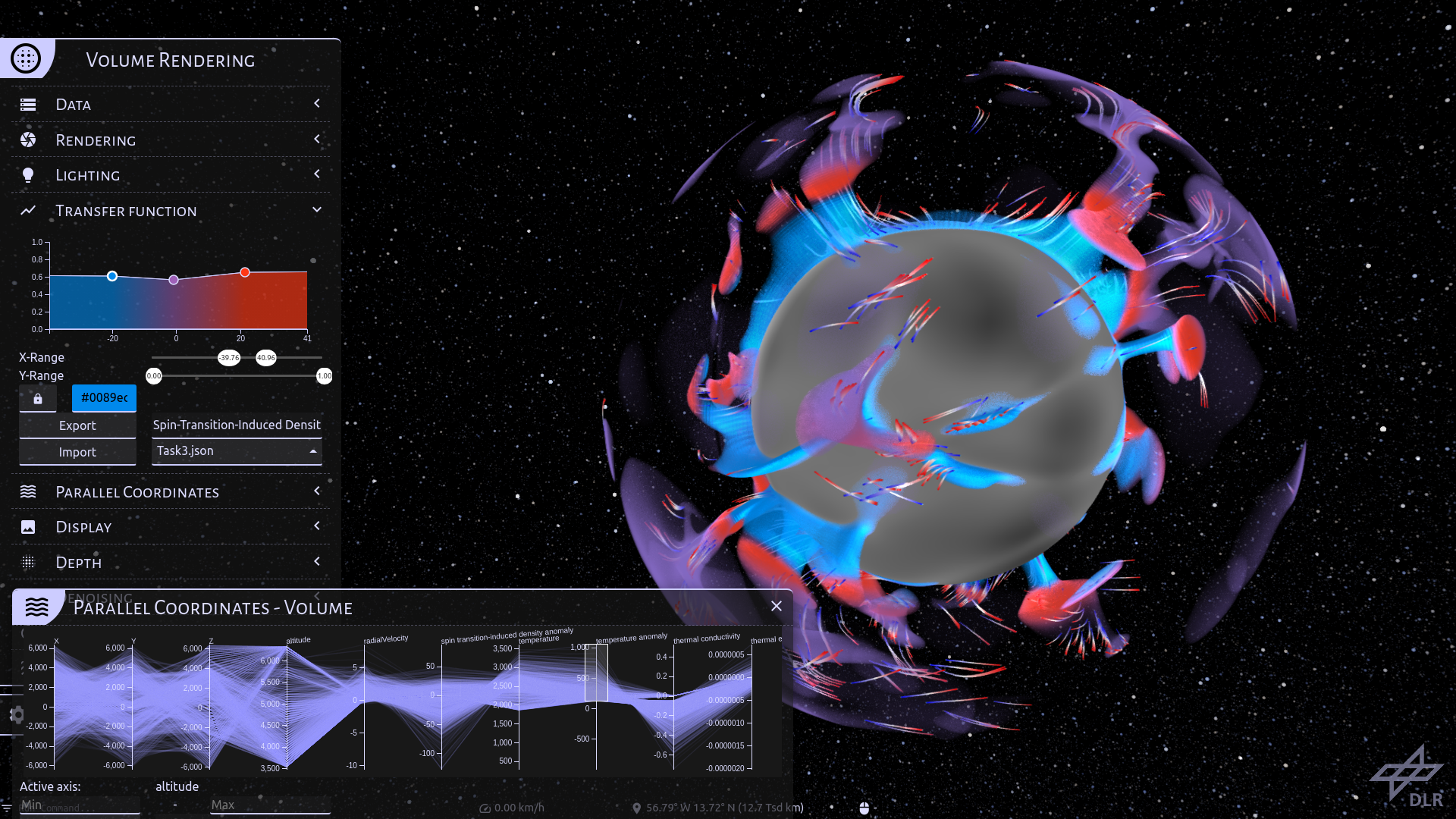}
    \caption{Pathlines and their color can be used to visualize the flow of material (blue pathline color: starting point of a fluid parcel, red pathline color: its location after 20\,Myrs).}
\end{figure*}

\begin{figure*}
    \centering
    \captionsetup{width=0.9\textwidth}
    \includegraphics[width=0.9\linewidth]{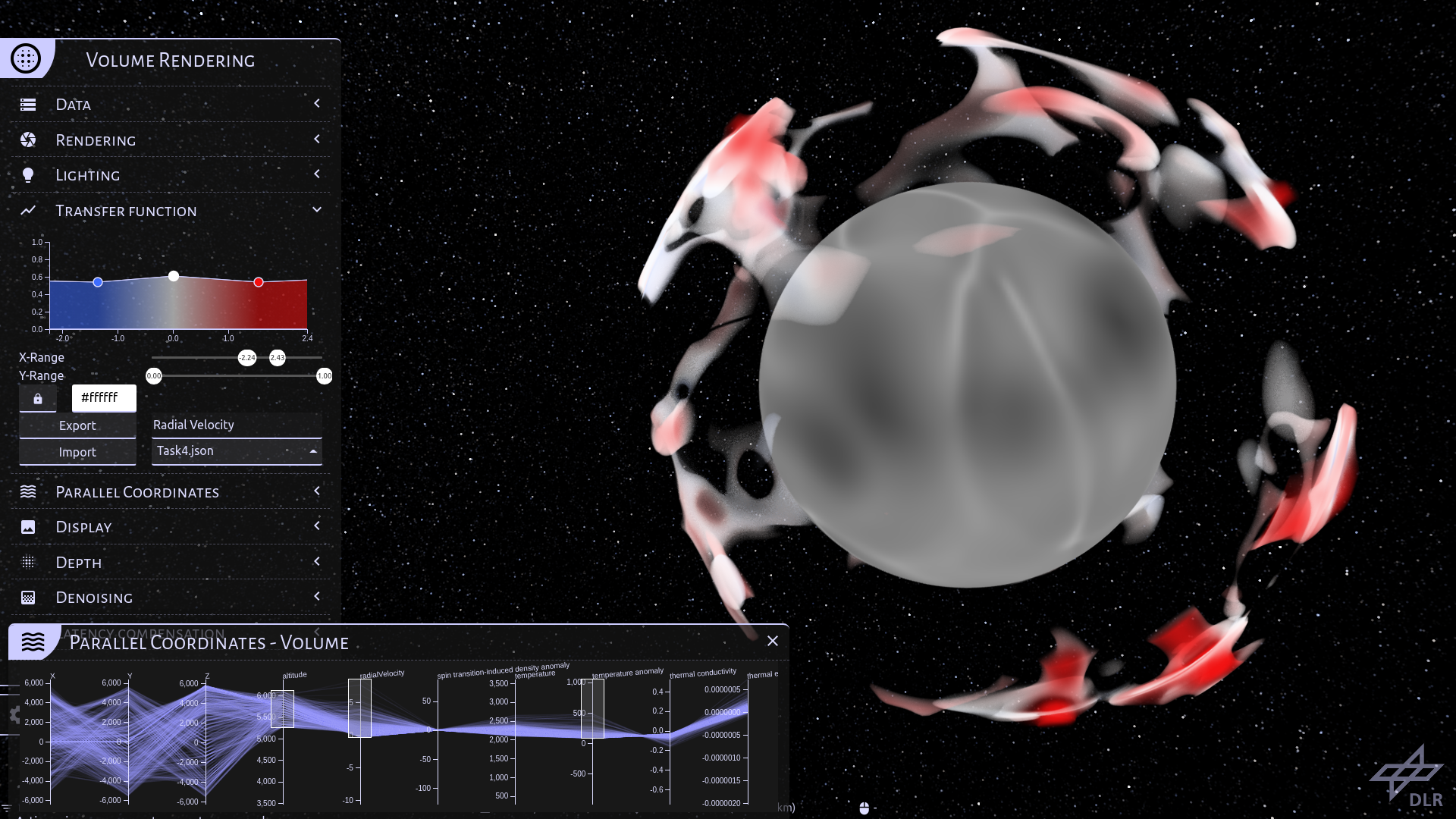}
    \caption{For \textbf{Task 4}, we selected a depth around 660\,km and a positive temperature anomaly. The transfer function uses the radial velocity. This causes rising material to be shown in red color while stagnating material is shown in white. In this time step there are several good examples of rising plumes with stagnated material at 660-phase boundary (at about two, four, five, and six o'clock).}
\end{figure*}

\clearpage
\newpage

\begin{figure*}
    \centering
    \captionsetup{width=\textwidth}
    \includegraphics[width=\textwidth]{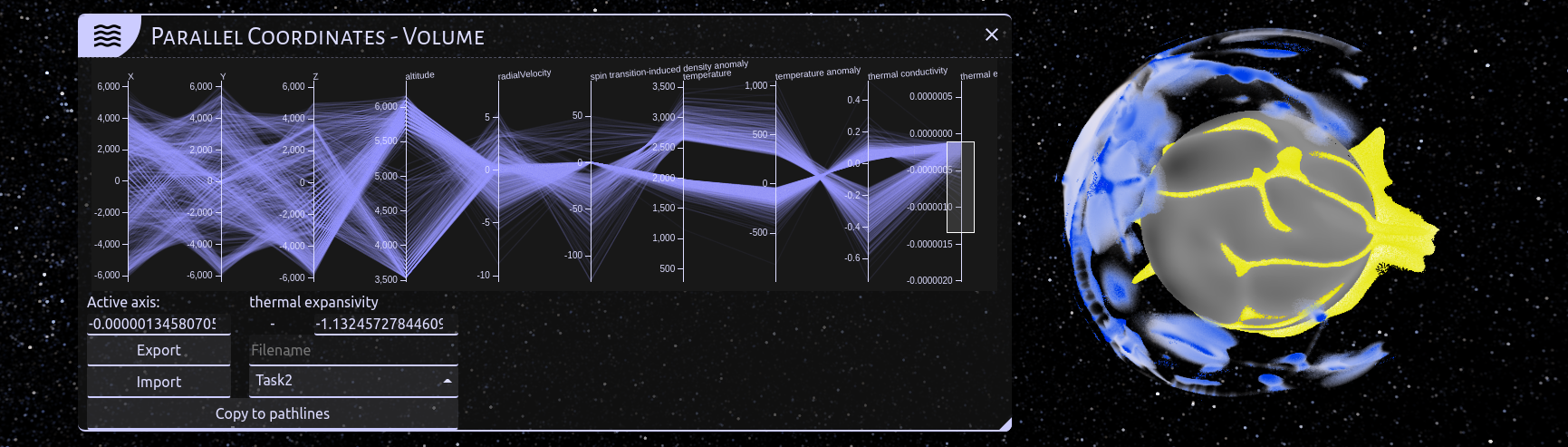}
    \caption{As an example for \textbf{Task 5}, we select a negative thermal expansivity. The lines of the parallel coordinates plot clearly show a separation of the data into two distinct regions. We can use further selections in the parallel coordinates to choose the individual regions. The transfer function uses the temperature anomaly.}
\end{figure*}

\begin{figure*}
    \centering
    \captionsetup{width=\textwidth}
    \includegraphics[width=\textwidth]{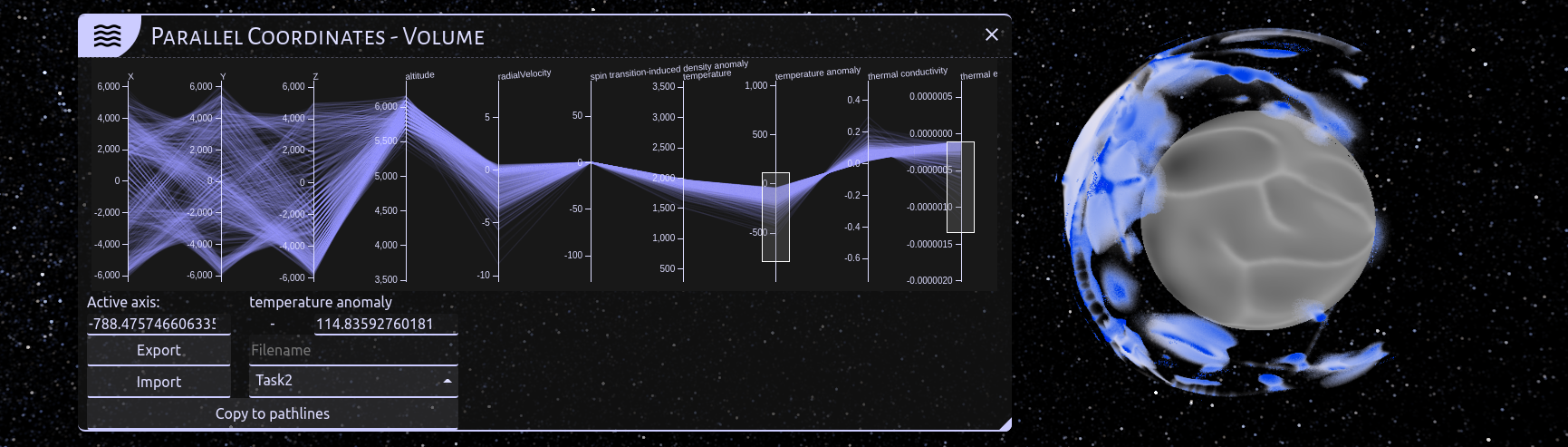}
    \caption{The relatively cold region at high altitudes.}
\end{figure*}

\begin{figure*}
    \centering
    \captionsetup{width=\textwidth}
    \includegraphics[width=\textwidth]{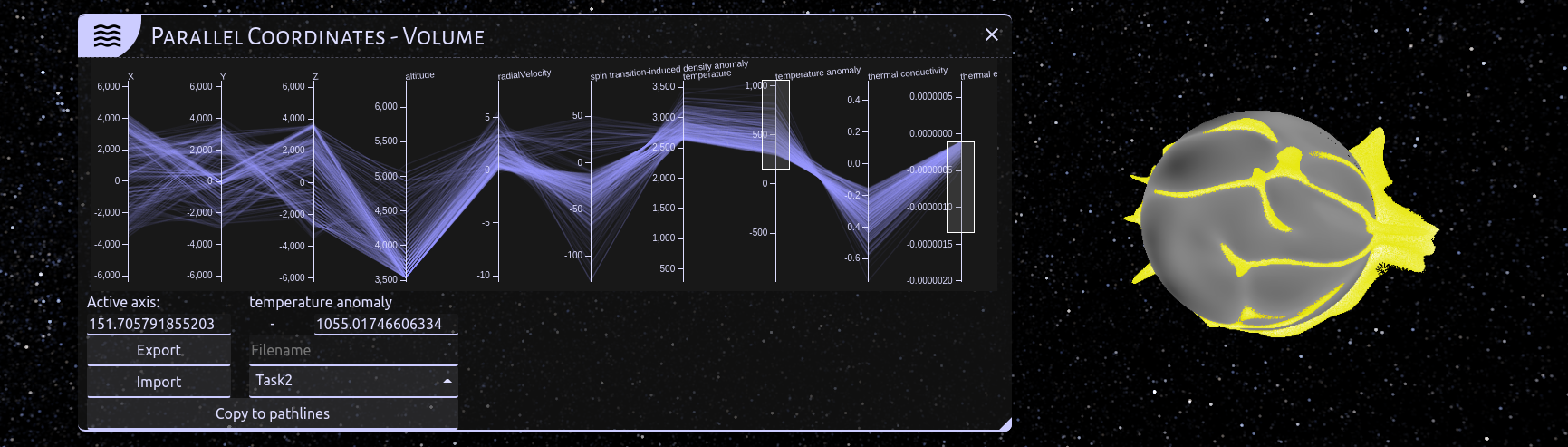}
    \caption{The relatively hot region at low altitudes.}
\end{figure*}

\begin{figure*}
    \centering
    \captionsetup{width=\textwidth}
    \includegraphics[width=\textwidth]{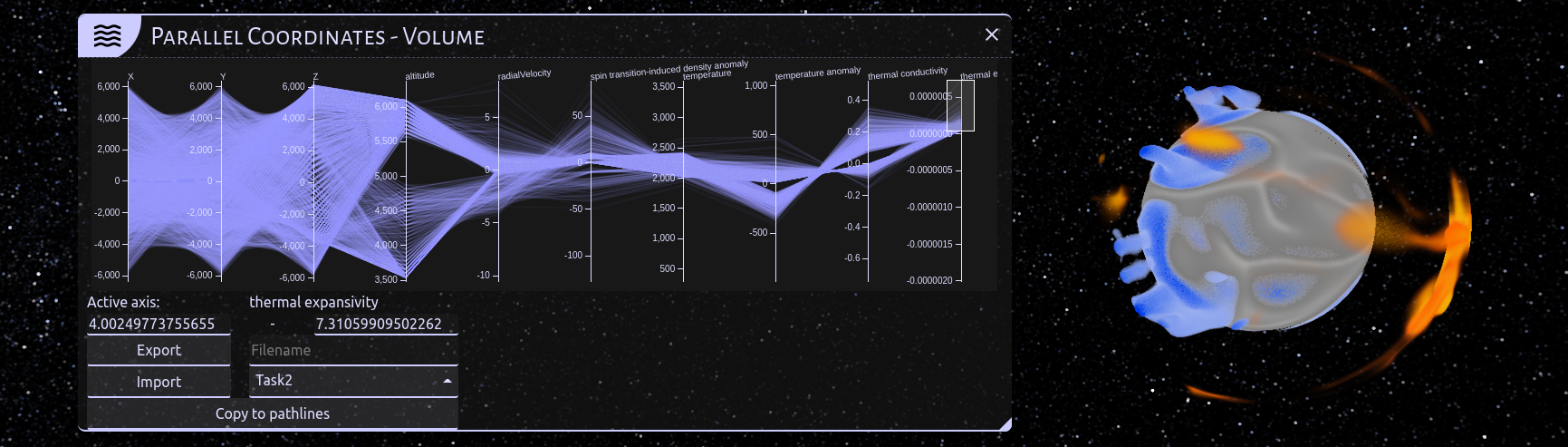}
    \caption{If we select a positive thermal expansivity, the data is also separated into two distinct regions. We can use further selections in the parallel coordinates to choose the individual regions. The transfer function again uses the temperature anomaly.}
\end{figure*}

\begin{figure*}
    \centering
    \captionsetup{width=\textwidth}
    \includegraphics[width=\textwidth]{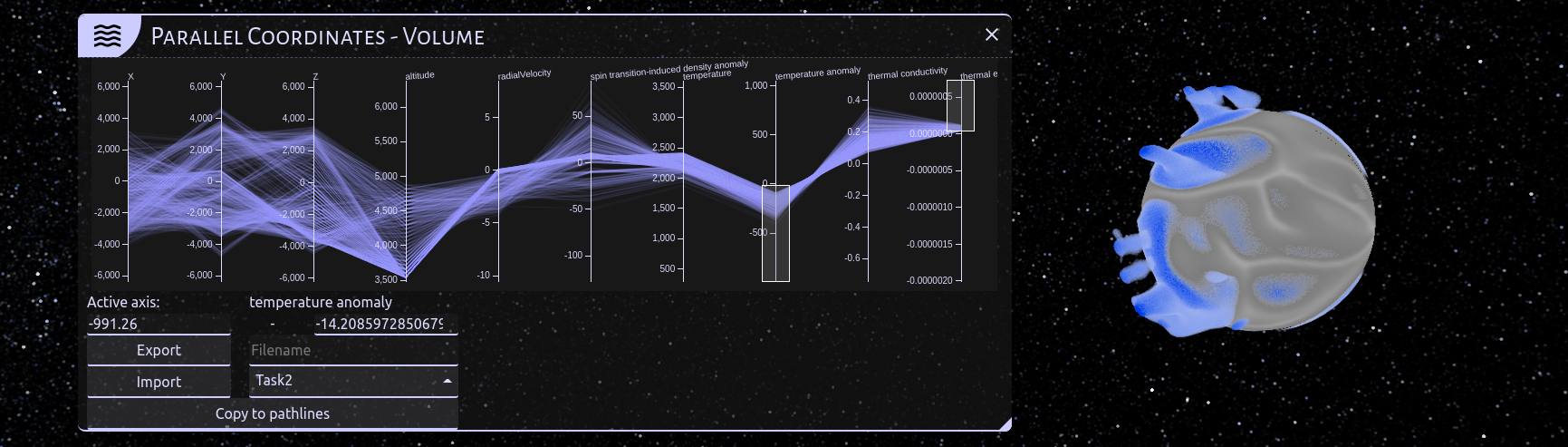}
    \caption{The relatively cold region at low altitudes.}
\end{figure*}

\begin{figure*}
    \centering
    \captionsetup{width=\textwidth}
    \includegraphics[width=\textwidth]{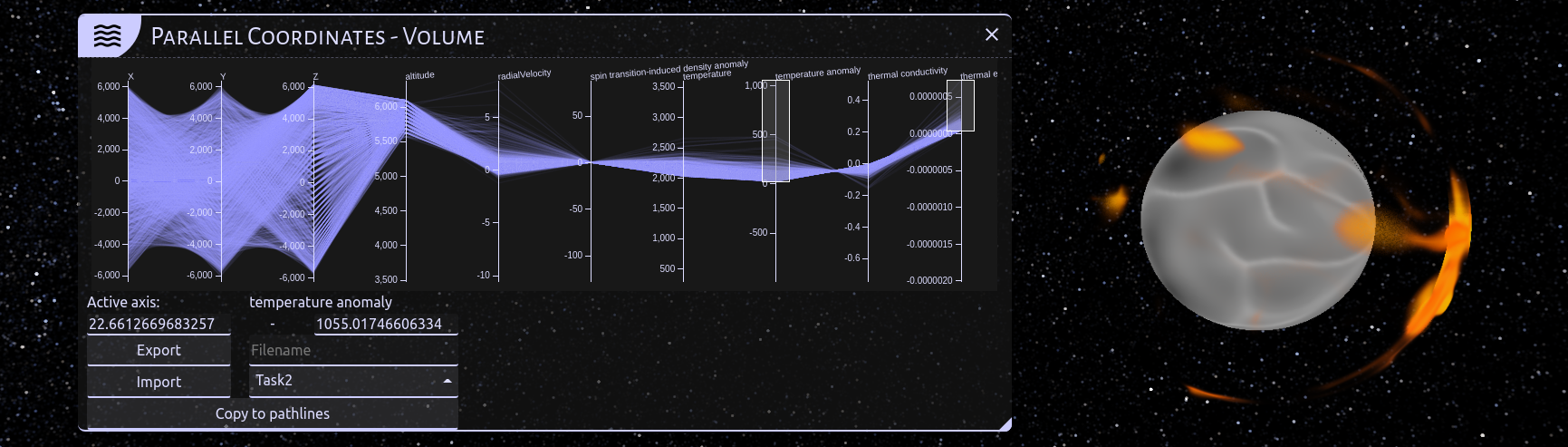}
    \caption{The relatively hot region at high altitudes.}
\end{figure*}

\end{document}